\documentstyle[preprint,version2,aps]{revtex}
\begin{document}
\begin{title}
Nonperturbative aspects of the quark-photon vertex
\end{title}
\author{M.R. Frank}
\begin{instit}
Institute for Nuclear Theory, University of Washington, HN-12, Seattle, WA
98195
\end{instit}
\begin{abstract}
The electromagnetic interaction with quarks is investigated through a
relativistic, electromagnetic gauge-invariant treatment. Gluon dressing
of the quark-photon vertex and the quark self-energy functions is
described by the inhomogeneous Bethe-Salpeter equation in the ladder
approximation and the Schwinger-Dyson equation in the rainbow
approximation respectively.  Results for the calculation of the
quark-photon vertex are presented in both the time-like and space-like
regions of photon momentum squared, however emphasis is placed on the
space-like region relevant to electron scattering.  The treatment
presented here simultaneously addresses the role of dynamically
generated $q\bar{q}$ vector bound states and the approach to asymptotic
behavior.  The resulting description is therefore applicable over the
entire range of momentum transfers available in electron scattering
experiments.  Input parameters are limited to the model gluon two-point
function, which is chosen to reflect confinement and asymptotic
freedom, and are largely constrained by the obtained bound-state spectrum.
\end{abstract}

\section{Introduction}
As hadrons are subjected to more detailed examination in electron
scattering experiments, such as those proposed at CEBAF, a quark-based
description of their electromagnetic (EM) interaction, which is
applicable over a broad range of momenta, becomes desirable.  Local
effective field theories based on hadronic degrees of freedom, although
attractive for their efficiency, are subject to difficulties when
conditions are sufficient to probe characteristics of subhadronic
origin\cite{isgurjaffe}.  An ideal perspective on this problem can be
gained by sacrificing local interactions in favor of retaining some
knowledge of the quark substructure content of the effective hadron
fields\cite{frank94}, and is the spirit of the present investigation.
In such a treatment, the EM interaction with a pion at tree level in
the hadron fields is, for example, illustrated in Fig.\ref{fig1}.
There, the pion Bethe-Salpeter amplitudes, $\Gamma $, reflect the
ability of the internal degrees of freedom to share the transferred
momentum, while the quark-photon vertex (QPV), $\Gamma _{\nu }$,
contains information about the interaction with a dressed quark,
including the production of $q\bar{q}$ vector-meson modes expected from
vector-meson dominance or dispersion theory phenomenology. At larger
momentum transfers where the effective scale provided by the
vector-meson poles has been exceeded, the QPV also provides information
about the approach of photon-hadron vertices  to asymptotic behavior.
A quantitative understanding of the QPV is therefore central to the EM
description of hadrons at intermediate momentum scales, and must
address issues of confinement, dynamically generated $q\bar{q}$ vector
bound states, and asymptotic freedom.

In addition to its role in the EM interaction, the QPV is a quantity of
interest on its own merit. It is well known, for example, that a formal
description of meson bound-state spectra and wavefunctions can be
obtained by considering the interaction of a quark with an external
field having the quantum numbers of the bound state of interest. In
such a description, the external field excites a $q\bar{q}$ pair from
the vacuum.  Their subsequent gluon-mediated interactions provide
self-energy and vertex dressing.  The existence of bound states is
exhibited by poles in the dressed vertex.  The position and residue of
a given pole yield  information about the bound-state mass and
wavefunction respectively.  Vertex functions of quarks with external
fields thus provide a unique environment in which to study the
formation of  meson bound states.  The interaction with a photon is
simply one example of this, and it is expected that the approach
presented here can be applied to the description of bound states with
other quantum numbers as well as vectors.

In this paper the QPV with gluon dressing  at the ladder level of
approximation is investigated in both the time-like and space-like
region of photon momentum squared.  Particular attention is given to
the formation of vector bound states and confinement phenomenology,
however the primary goal of this initial investigation is to develop a
description of the quark-photon interaction in the space-like region
relevant to electron scattering which simultaneously addresses the role
of vector mesons and quark substructure.  The equations implied by the
ladder approximation are$^1$ the inhomogeneous Bethe-Salpeter equation
\begin{equation}
\Gamma _{\mu }(P,q)=-i\gamma _{\mu }-\frac{4}{3}g^2\int
\frac{d^4K}{(2\pi )^4}D(P-K)\gamma _{\nu }G(K_{+})\Gamma _{\mu
}(K,q)G(K_{-})\gamma _{\nu },\label{IBS}
\end{equation}
and the Schwinger-Dyson equation
\begin{equation}
\Sigma (P)=\frac{4}{3}g^2\int \frac{d^4K}{(2\pi )^4}D(P-K)\gamma _{\nu
}G(K)\gamma _{\nu },\label{SDE}
\end{equation}
where $K_{\pm }=K\pm q/2$, the quark Green's function, $G$, is defined
from its inverse \mbox{$G^{-1}(P)=i\! \not \! P+\Sigma (P)$}, and for
convenience the model gluon two-point function is taken to be diagonal
in Lorentz indices ($D_{\mu \nu}=\delta _{\mu \nu }D$).  In
Eq.(\ref{IBS}) $q$ is the photon momentum, while $P$ is the average of
the incoming and outgoing quark momenta.
These expressions are illustrated in Fig.\ref{fig2}.  The current-quark
masses in this investigation are assumed to be zero, which allows an
overall charge matrix to be omitted from Eq.(\ref{IBS}), but the
approach can be easily extended to include nonzero masses.  The absence
of masses further implies that all dimensionful constants enter through
the parameterization of the gluon two-point function, $D$.  Here the
infrared contribution to this quantity is characterized by an effective
strength, $\eta $, and an effective range, $R_0$, in addition to the
known ultraviolet form.

The motivation for employing the ladder approximation is provided by
the following arguments.  First, it offers an EM gauge-invariant
description.  Both the Ward-Takahashi identity (WTI),
\begin{equation}
q_{\mu }\Gamma _{\mu }(P,q)=G^{-1}(P_{-})-G^{-1}(P_{+}),\label{WTI}
\end{equation}
and the Ward identity (WI),
\begin{equation}
\Gamma _{\mu }(P,0)=-\frac{\partial G^{-1}(P)}{\partial P_{\mu }}, \label{WI}
\end{equation}
are satisfied by this approximation.  This is easily demonstrated by
directly substituting (\ref{WTI}) or (\ref{WI}) into (\ref{IBS}) for
the vertex, and
employing (\ref{SDE}).  Further, it has been shown recently how gauge
invariance at the quark level in the ladder approximation to the QPV
fits into a gauge-invariant description of composite mesons leading to
WTIs at the hadronic level\cite{frank94}.

Second, in the vicinity of a $q\bar{q}$ resonance, the vertex obtains the form
\begin{equation}
\Gamma _{\mu }(P,q) \approx \frac{M_V^2}{q^{2}+M_{V}^{2}}\Omega _{\mu
}(P,q),\label{pole}
\end{equation}
where $\Omega _{\mu }$ is the residue at the pole, and from (\ref{IBS})
satisfies
\begin{equation}
\Omega _{\mu }(P,q)=-\frac{4}{3}g^2\int \frac{d^4K}{(2\pi
)^4}D(P-K)\gamma _{\nu }G(K_{+})\Omega _{\mu }(K,q)G(K_{-})\gamma _{\nu
},\label{BS}
\end{equation}
for $q^2=-M^2_V$.  This is the homogeneous Bethe-Salpeter equation in
the ladder approximation for a vector $q\bar{q}$ bound state.  One can
conclude from (\ref{WTI}) and (\ref{pole}) that the on-mass-shell
solution of the homogeneous equation satisfies $q_{\mu }\Omega _{\mu
}(P,q)=0$, provided the self-energy functions are nonsingular.  The
Bethe-Salpeter equation at this level has been investigated
previously, and acceptable results have been obtained for the lowest
mass bound states\cite{CRP87,PCR89,aoki91,MJ92}.  It is therefore
anticipated that, in the addition to providing an EM gauge invariant
description of the quark-photon interaction for use in electron
scattering, this level of approximation has the ability to describe the
spectrum and structure of vector bound states.

Finally, the Schwinger-Dyson equation (\ref{SDE}) has been extensively
studied with regard to confinement phenomenology and dynamical chiral
symmetry breaking\cite{conchi}.  It has been demonstrated, for example,
that for a suitably chosen gluon two-point function $D$, quark Green's
functions which are free of singularities on the real $P^2$ axis are
produced; therefore denying the existence of a single-particle
spectrum.  It is shown here, however, that this is not a sufficient
condition for the formation of finite-size $q\bar{q}$ states, but is an
essential ingredient in a description of confinement based on an
interaction between dressed quarks which does not rise to infinity at
large distances.  This alternate description of confinement in fact
relies on the interaction between dressed quarks diminishing at
separations beyond the characteristic length scale, $R_0$, at which
point the quark Green's function approaches its vacuum value.  The
quarks are then repelled by the vacuum due to absence of a mass pole in
the vacuum quark Green's function.  The effective ``rising potential"
of traditional models of confinement occurs here as a result of the
repulsive interaction of quarks with the vacuum in combination with the
decreasing  interaction between dressed quarks with increasing
separation. Such a description has proven successful in the formulation
of a confining nontopological soliton model for baryons\cite{GSM}.
This mechanism arises naturally here, and is pursued as a possible
realization of confinement.

The paper is organized as follows.  In  Sec.2 the general features of
the QPV are given along with the reduction of Eq.(\ref{IBS}) to a form
suitable for computer applications.  The solution of the homogeneous
Bethe-Salpeter equation (\ref{BS}) is also discussed there.  In Sec.3 a
discussion of confinement is presented in the context of a particular
model which yields analytic solutions to both the inhomogeneous
Bethe-Salpeter and Schwinger-Dyson equations (\ref{IBS}) and
(\ref{SDE}) respectively.  Also discussed there is the criteria for the
formation of finite-range bound states in the context of this model of
confinement.  The numerical evaluation of the QPV for both the
time-like and space-like regions of the photon momentum squared is
given in Sec.4.  Finally, a summary is presented in Sec.5.

\section{General features}
\subsection{The inhomogeneous Bethe-Salpeter equation}

In this section, the steps required to reduce the inhomogeneous
Bethe-Salpeter equation (\ref{IBS}) to a form suitable for computer
applications are outlined.  The approach is straightforward and is
based on the general features of the QPV as described by (\ref{IBS}).
Where possible, lengthy algebraic manipulations are omitted.  To begin,
it is useful to write the self energy, $\Sigma $ from (\ref{SDE}), in the form
\begin{equation}
\Sigma (P)=i\! \not \! P\left[ A(P^2)-1\right] +B(P^2)\label{2.1}
\end{equation}
where
\begin{equation}
\left[ A(P^2)-1\right] P^2= g^2\frac{8}{3}\int \frac{d^4K}{(2\pi
)^4}D(P-K)\frac{A(K^2)P\cdot K}{K^2A^2(K^2)+B^2(K^2)},\label{2.2}
\end{equation}
and
\begin{equation}
B(P^2)=g^2\frac{16}{3}\int \frac{d^4K}{(2\pi
)^4}D(P-K)\frac{B(K^2)}{K^2A^2(K^2)+B^2(K^2)}.\label{2.3}
\end{equation}
The quark Green's function, $G$, and its inverse, $G^{-1}$, can then be written
as
\begin{equation}
G(P)=-i\! \not \! P \alpha (P^2)+\beta (P^2)\; \; \; \; \; \mbox{and}
\; \; \; \; \;G^{-1}(P)=i\! \not \! PA(P^2)+B(P^2),\label{2.4}
\end{equation}
where $\alpha $ and $\beta $ are defined in an obvious fashion from $G^{-1}$.

{}From Eq.(\ref{IBS}) it is evident that the QPV, in addition to its
Lorentz four-vector structure, is a four-by-four matrix in Dirac
indices and is in general a function of two non-orthogonal four
vectors, $P$ and $q$ ($P\cdot q \neq 0$).  On first appraisal, one
might therefore expect that the complete set of sixteen Dirac matrices
$\left\{{\bf1},\gamma _5,\gamma _{\mu },\gamma _5\gamma _{\mu },\sigma
_{\mu \nu}\right\}$ must be employed in its description.  However, by
limiting the gluon two-point function, $D$, to diagonal components in
Lorentz indices, the tensor, $\sigma _{\mu \nu }$, is excluded.  This
is seen explicitly by rewriting the integrand of (\ref{IBS}) using
\begin{eqnarray}
\left[ \gamma _{\nu }G(K_+)\Gamma _{\mu }(P,q)G(K_-)\gamma _{\nu
}\right] _{il}&=&\left( \gamma _{\nu }\right) _{ij}\left[ G(K_+)\Gamma
_{\mu }(P,q)G(K_-)\right] _{jk}\left( \gamma _{\nu }\right) _{kl}\nonumber \\
&=&\frac{1}{4}M^a_{il}\mbox{tr}\left[ M^aG(K_+)\Gamma _{\mu
}(P,q)G(K_-)\right] ,\label{2.5}
\end{eqnarray}
where the set of matrices $M^a\equiv \left\{ {\bf1},i\gamma _5,i\gamma
_{\mu }/\sqrt{2},i\gamma _5\gamma _{\mu }/\sqrt{2}\right\} $, from
which $\sigma _{\mu \nu }$ is absent, are obtained through Fierz
reordering, and the repeated indices are summed.  Further, the
pseudo-scalar matrix, $\gamma _5$, can be excluded by considering that
the corresponding term in (\ref{2.5}) has the form $\gamma _5\Lambda
^5_{\mu }(P,q)$, where the matrix structure has been completely
factored.  In order to contribute to the QPV, this term must form a
four vector, which requires $\Lambda ^5_{\mu }(P,q)$ to be an axial
vector.  It is immediately evident that this is impossible since there
is an insufficient number of four vectors ($P$ and $q$) to combine with
the tensor $\epsilon _{\mu \nu \alpha \beta }$ to form an axial vector.
 The solution to (\ref{IBS}) must therefore have the general form
\begin{equation}
\Gamma _{\mu }(P,q)={\bf 1}\Lambda ^{(1)}_{\mu }(P,q)+i\gamma _{\nu
}\Lambda ^{(2)}_{\nu \mu }(P,q)+i\gamma _5\gamma _{\nu }\epsilon _{\mu
\nu \alpha \beta }P_{\alpha }q_{\beta }\eta ^{-2}\Lambda
^{(3)}(P,q),\label{2.6}
\end{equation}
where the matrix structure is now explicit and the quantities $\Lambda
^{(i)}$ are dimensionless functions with Lorentz structure indicated by
their indices.  The constant $\eta $ has dimension of (length)$^{-1}$
and is associated with the infrared strength of the gluon two-point function.

The further reduction of the $\Lambda ^{(i)}$ to a set of invariant
functions is achieved through the use of the symmetry transformations
of the vertex $\Gamma _{\mu }(P,q)$ under $\gamma _5$; charge
conjugation, ${\cal C}=\gamma _2\gamma _4$; and parity, ${\cal
P}=\gamma _4$.  The transformation properties are determined directly
from the inhomogeneous Bethe-Salpeter equation (\ref{IBS}) and are given by
\begin{eqnarray}
\gamma _5\Gamma _{\mu }(P,q)\gamma _5 &=& -\Gamma _{\mu }(-P,-q)\nonumber \\
{\cal C}\Gamma _{\mu }(P,q){\cal C}^{-1}&=& -\Gamma _{\mu
}(-P,q)^t\label{2.7}\\
{\cal P}\Gamma _{\mu }(P,q){\cal P}^{-1}&=& \omega _{\mu \nu }\Gamma
_{\nu }(P_4;-{\bf P},q_4;-{\bf q})\nonumber
\end{eqnarray}
where $\omega _{\mu \nu }=$diag$(-1,-1,-1,1)$ and $t$ denotes a matrix
transpose.  From the general form given in (\ref{2.6}) and the first
two expressions in (\ref{2.7}), it is clear that $\Lambda ^{(2)}_{\mu
\nu }(P,q)$ and $\Lambda ^{(3)}(P,q)$ are even in both $P$ and $q$,
while $\Lambda ^{(1)}_{\mu }(P,q)$ is odd in $P$ and even in $q$.  The
quantities $\Lambda ^{(i)}$ can therefore be written as
\begin{eqnarray}
\Lambda^{(1)}_{\mu }(P,q)&=&\frac{q_{\mu }}{\eta }\frac{P\cdot
q}{q^2}\lambda ^L_1+\frac{P^T_{\mu}}{\eta }\lambda ^T_1,\nonumber \\
\Lambda ^{(2)}_{\nu \mu }(P,q)&=&\frac{P_{\nu }q_{\mu }}{\eta
^2}\frac{P\cdot q}{q^2}\lambda ^L_2+\frac{P_{\nu }P^T_{\mu }}{\eta
^2}\lambda ^T_2\label{2.8}\\
& &-\frac{q_{\nu }q_{\mu }}{q^2}\lambda ^L_3-\left( \delta _{\nu \mu
}-\frac{q_{\nu }q_{\mu }}{q^2}\right) \lambda ^T_3 + \frac{q_{\nu
}P^T_{\mu }P\cdot q}{\eta ^4}\lambda ^T_4,\nonumber \\
\Lambda ^{(3)}(P,q)&=&\lambda ^T_5.\nonumber
\end{eqnarray}
Here the coefficients, $\lambda $, are functions of $P^2$, $q^2$, and
$C^2_{Pq}$ where $C_{Pq}\equiv P\cdot q/(Pq)$ is the direction cosine
between $P$ and $q$, and $P^T_{\mu }\equiv P_{\mu}-P\cdot qq_{\mu
}/q^2$ is the vector transverse to $q_{\mu }$ ($P^T\cdot q=0$).  The
behavior of the coefficients for large space-like $q^2$ consistent with
asymptotic freedom\cite{marciano78} can be anticipated from (\ref{2.6})
and (\ref{2.8}).  In particular, as $q^2\rightarrow \infty $ the
coefficients $\lambda ^{T,L}_3$ must tend to unity, while all others tend to
zero.
The separation in (\ref{2.8}) into longitudinal and transverse
contributions with respect to the index $\mu $ allows three of the
eight coefficient functions, $\lambda $, to be immediately determined
from the WTI (\ref{WTI}) in terms of the self-energy functions $A$ and
$B$ given in (\ref{2.2}) and (\ref{2.3}) respectively.  The
longitudinal coefficient functions are thus given by
\begin{eqnarray}
\frac{P\cdot q}{\eta }\lambda ^L_1&=&B(P^2_-)-B(P^2_+)\nonumber \\
\frac{P\cdot q}{\eta ^2}\lambda ^L_2&=&A(P^2_-)-A(P^2_+)\label{2.9} \\
\lambda ^L_3&=&\frac{1}{2}\left[ A(P^2_-)+A(P^2_+)\right]  .\nonumber
\end{eqnarray}
The additional constraint provided by the WI (\ref{WI}) obtains the
soft-photon limits:
\begin{eqnarray}
\left. \lambda ^L_1\right| _{q=0}=\left. \lambda ^T_1\right|
_{q=0}&=&-2\eta B'(P^2)\nonumber \\
\left. \lambda ^L_2\right| _{q=0}=\left. \lambda ^T_2\right|
_{q=0}&=&-2\eta ^2A'(P^2)\label{2.10}\\
\left. \lambda ^L_3\right| _{q=0}=\left. \lambda ^T_3\right|
_{q=0}&=&A(P^2),\nonumber
\end{eqnarray}
where the prime denotes differentiation with respect to the argument.

The five unknown transverse coefficients can be isolated using
(\ref{2.6}) and (\ref{2.8}), and the orthogonality of the Dirac
matrices.  The closed set of equations so obtained can be written in matrix
form as
\begin{eqnarray}
b_i&=&{\cal N}_{ij}(P^2,C_{Pq}^2;q^2)\lambda ^T_j(P^2,C_{Pq}^2;q^2)\nonumber \\
& & \; \; \; \; \; \; + \int _{0}^{\infty }dK K^3\int
_{-1}^{1}dC_{Kq}{\cal M}_{ij}(P^2,C_{Pq}^2,K^2,C_{Kq}^2;q^2)\lambda
^T_j(K^2,C_{Kq}^2;q^2),\label{2.16}
\end{eqnarray}
where ${\cal M}$ and ${\cal N}$($b$ and $\lambda ^T$) are
matrices(vectors) in the five-component space of the coefficient
functions and $C_{Kq}$ is the direction cosine between $K$ and $q$.
The steps required to reach the result (\ref{2.16}) along with the
explicit forms of ${\cal M}$, ${\cal N}$ and $b$ are given in the
appendix.  The equation (\ref{2.16}) contains the same information as
the transverse component of the inhomogeneous Bethe-Salpeter equation
(\ref{IBS}); to this point, no approximations have been made.  To
proceed further requires the specification of the gluon two-point function,
$D$.

\subsection{The homogeneous Bethe-Salpeter equation}

The solutions of the homogeneous Bethe-Salpeter equation, (\ref{BS}),
describe the coupling of a $q\bar{q}$ vector bound state to a quark,
and thereby provide information about the internal dynamics of the
bound state.  Although these solutions can be obtained directly from
the coefficient functions, $\lambda ^{T}_{i}$ of the previous section,
their normalization is more readily achieved  by considering the
solution of the homogeneous equation.  To this end, the quantity
$\Omega _{\mu }(P,q)$ is decomposed as
\begin{equation}
\Omega _{\mu }(P,q)={\bf 1}\phi ^{(1)}_{\mu
}(P,q)+\frac{i}{\sqrt{2}}\gamma _{\nu }\phi ^{(2)}_{\nu \mu
}(P,q)+\frac{i}{\sqrt{2}}\gamma _{5}\gamma _{\nu }\phi ^{(3)}_{\nu \mu
}(P,q).\label{2.17}
\end{equation}
The association of the coefficients $\phi ^{(i)}$ with the coefficients
 $\Lambda ^{(i)}$ of (\ref{2.6}) is directly observed from the
relationship between the solutions to the inhomogeneous and homogeneous
Bethe-Salpeter equations given in (\ref{pole}).  The substitution of
(\ref{2.17}) into the homogeneous equation (\ref{BS}) allows a
description in the form of the eigenvalue equation
\begin{equation}
\int d^4K\Delta ^{-1}_{ab}(P,K;q)\Phi _{b}^{\nu }(K,q)=\alpha (q^2)\Phi
_{a}^{\nu }(P,q)\label{2.18}
\end{equation}
to be obtained. The quantity $\Delta ^{-1}_{ab}(P,K;q)$ is defined as
\begin{equation}
\Delta ^{-1}_{ab}(P,K;q)\equiv \frac{9}{2}\int \frac{d^4r}{(2\pi
)^4}\frac{e^{-i(P-K)r}}{g^2D(r)}\delta _{ab}
+\delta ^{(4)}(P-K)\mbox{tr}\left[ M^aG(K_{+})M^bG(K_{-})\right] \label{2.19}
\end{equation}
and plays the role of an inverse propagator for the $q\bar{q}$
composite system, as has been demonstrated in previous work on the
hadronization of quark field-theory models\cite{localization}.  That
description is extended here to include the more general structure in
(\ref{2.17}).  The eigenfunctions are defined in terms of the
coefficient functions, $\phi ^{(i)}$, as $\Phi _{1}^{\nu }\equiv \phi
^{(1)}_{\nu }$, $\Phi _{2-5}^{\nu }\equiv \phi ^{(2)}_{\alpha \nu }$,
and $\Phi _{6-9}^{\nu }\equiv \phi ^{(3)}_{\alpha \nu }$.  The matrices
$M^a\equiv {\bf 1}_C{\bf 1}_F\left\{ {\bf 1},i\gamma _{\mu
}/\sqrt{2},i\gamma _{5}\gamma _{\mu }/\sqrt{2}\right\} $ include unity
in the color and two-component flavor space, and the trace is taken to
include their indices as well as those of the Dirac matrices.  The
homogeneous equation (\ref{BS}) is regained through the requirement
that the eigenvalue vanishes on the mass shell, that is, $\alpha
(q^2=-M_V^2)=0$.  The normalization condition for the eigenfunctions is
obtained by multiplying (\ref{2.17}) on the left by $\Phi _{a}^{\mu
}(P,-q)$ and integrating over relative momentum $P$, which gives
\begin{eqnarray}
\int d^4Pd^4K\Phi _{a}^{\mu }(P,-q)\Delta ^{-1}_{ab}(P,K;q)\Phi
_{b}^{\nu }(K,q)&=&\alpha (q^2)\int d^4P\Phi _{a}^{\mu }(P,-q)\Phi
_{a}^{\nu }(P,q)\nonumber \\
&\stackrel{q^2\sim -M_V^2}{=} & \left(\delta _{\mu \nu}-\frac{q_{\mu
}q_{\nu }}{q^2}\right) \left(q^2+M_V^2\right) Z(q^2).\label{2.20}
\end{eqnarray}
The last line in (\ref{2.19}) follows from the fact that the
on-mass-shell  eigenfunctions, $\Phi _{a}^{\mu }$, contain only
components transverse to the center-of-mass momentum $q_{\mu }$.  The
(dimensionless) wavefunction renormalization constant $Z(q^2)$ can be
absorbed into the Bethe-Salpeter amplitudes by defining $\hat{\Phi
}_{a}^{\nu }\equiv \Phi _{a}^{\nu }/\sqrt{Z(-M_V^2)}$.  The
normalization condition can then be written explicitly as
\begin{equation}
\left. \frac{1}{3}\int d^4Pd^4K\hat{\Phi }_{a}^{\nu
}(P,-q)\frac{\partial }{\partial q_{\mu }}\left[ \Delta
^{-1}_{ab}(P,K;q)\right] \hat{\Phi }_{b}^{\nu }(K,q)\right| _{q^2=-M_V^2}
=2q_{\mu },\label{2.21}
\end{equation}
which is the standard result\cite{iz}.  In terms of the properly
normalized Bethe-Salpeter amplitude, the relationship to the QPV given
in (\ref{pole}) now reads
\begin{equation}
\Gamma _{\mu }(P,q) \approx
\frac{M_V^2}{q^{2}+M_{V}^{2}}\frac{\hat{\Omega }_{\mu }(P,q)}{f_V},\label{2.22}
\end{equation}
where $f_V\equiv 1/\sqrt{Z(-M_V^2)}$ is the effective coupling of the
vector meson to the photon.

\section{A simple model}

Traditional descriptions of quark confinement are based on a potential
between quarks, in an appropriate color combination, which rises with
increasing separation (to infinity in the absence of pair creation).
Explored here is the notion that this effect may be in part due to the
interaction of quarks (or more generally colored objects) with the
vacuum.  In particular, that confinement is manifest in the absence of
a mass pole in the vacuum Green's function of a colored object implies
that the object is repelled by the vacuum, and hence attracted to other
colored objects to form a color singlet by virtue of the interaction
with the vacuum.  An illustration of this mechanism is afforded through
the use of a simple model\cite{gluon,GSM} of the gluon two-point function given
by
\begin{equation}
D\left( (P-K)^2\right) = \frac{3\eta ^2\pi ^4}{g^2}\delta
^{(4)}(P-K).\label{3.1}
\end{equation}
At the present level of approximation, the self-energy dressing from
the vacuum is provided by the equations (\ref{2.2}) and (\ref{2.3}).
The solutions to these implied by (\ref{3.1}) are
\begin{eqnarray}
A(P^{2}) & = & \left\{
\begin{array}{cl}
2, & P^{2}\leq \frac{\eta ^{2}}{4} \\
\frac{1}{2}\left[ 1+\left(1+\frac{2\eta ^{2}}{P^{2}}\right)
^{\frac{1}{2}}\right] , & P^{2}\geq \frac{\eta ^{2}}{4}
\end{array}\right. \nonumber \\ & &  \nonumber \\
B(P^{2}) & = & \left\{
\begin{array}{cl}
(\eta ^{2}-4P^{2})^{\frac{1}{2}}, &
\; \; \; \; \; \; \; \; \; \; \; \; P^{2}\leq \frac{\eta ^{2}}{4} \\
0, & \; \; \; \; \; \; \; \; \; \; \; \; P^{2}\geq \frac{\eta ^{2}}{4}
\end{array}\right. . \label{3.2}
\end{eqnarray}
That the solutions (\ref{3.2}) produce a model of confinement as
described above can be seen by the absence of a solution to the
expression $P^2+M^2(P^2)=0$, with $M=B/A$.  The presence of other
colored objects in a color singlet configuration can supply the
additional interaction necessary for the formation of a propagating
mode, as is demonstrated in the nontopological soliton model of
Ref.\cite{GSM}.  There, for example, in the case of quarks coupled to a
{\it constant} scalar mean field by the self-energy function $B$, the
quark inverse Green's function obtains the form $G^{-1}(P)=i\! \not \!
PA(P^2)+B(P^2)(1+\chi )$, where $0\geq \chi \geq -2$ characterizes the
strength of the mean field.  In this case a {\it continuous}
single-particle energy spectrum, $E^2({\bf P}^2)={\bf P}^2+M_C^2$, is
obtained because of the infinite-range potential, with the constituent
mass given by
\begin{equation}
M_C^2=\frac{\eta ^2}{4}\frac{(1+\chi )^2}{1-(1+\chi )^2}.\label{3.3}
\end{equation}
{}From (\ref{3.3}) it is evident that the increase of the constituent
mass as the strength, $\chi $, of the mean field decreases toward its
vacuum value ($\chi =0$), is due to the repulsive interaction with the
vacuum.  For the case of the constant mean field, the quarks are
allowed to propagate throughout space.  However, due to the energy
stored in the mean field, the self-consistent (minimum-energy) solution
acquires a finite range. In that case, as a quark in the system is
separated from the others, the influence of the mean field on the quark
diminishes and the mass rises as in (\ref{3.3}).  The quarks are thus
confined to the region of nonzero mean field by virtue of their
interaction with the vacuum.

The situation is similar for the QPV of interest here.  The use of the
gluon two-point function (\ref{3.1}) reduces the equations in
(\ref{2.16}) to algebraic form, which can then be solved by numerical
or symbolic techniques.  Shown in Fig.3 is the solutions for the five
transverse coefficient functions in both the time-like and space-like
regions of photon momentum $q^2$ for $P^2=C_{Pq}=0$.  The single
parameter, and the only dimensionful constant, $\eta $, is given the
value $5$ fm$^{-1}$.  It appears from Fig.3a that there is a pole for
$q^2=-\eta ^2/2$.
Closer analysis of the $q^2-P^2$ plane shows, however, that the
singularity structure is not a discrete pole, but is instead a
continuous spectrum.  For the region $P_{\pm}^2\leq \eta ^2/4$, the
analytic solutions have the simple form
\begin{eqnarray}
\lambda ^T_1&=&\frac{4\eta \left[ B(P^2_{+})+B(P^2_{-})\right]}{\eta
^2-4P^2+5q^2+B(P^2_{+})B(P^2_{-})}\nonumber \\
\lambda ^T_2&=&-\frac{q^2}{\eta ^2}\lambda ^T_4\nonumber \\
\lambda ^T_3&=&\frac{2\left[ 2\eta
^2-4P^2+q^2+B(P^2_{+})B(P^2_{-})\right] }{3\eta ^2-8P^2+6q^2}\label{3.4} \\
\lambda ^T_4&=&\frac{-32\eta ^4}{\left[ \eta
^2-4P^2+5q^2+B(P^2_{+})B(P^2_{-})\right] \left[ 3\eta
^2-8P^2+6q^2\right] }\nonumber \\
\lambda ^T_5&=&\frac{8\eta ^2}{3\eta ^2-8P^2+6q^2}\nonumber
\end{eqnarray}
from which the singularity structure is visible.  The self-energy
function $B$ is given in (\ref{3.2}).  The analytic solutions outside
this region are not as simple, but can be represented graphically.  As
an example, the coefficient function $\lambda ^T_3$ is shown in
Fig.\ref{fig4} for $C_{Pq}=0$.  The exhibited continuous spectrum
arises from the fact that the gluon two-point function (\ref{3.1})
provides a constant, or infinite-range, interaction between the quarks
in coordinate space.  The quarks are therefore allowed to propagate
throughout space uninhibited by the presence of the vacuum.  This
situation is analogous to that of the constant mean-field case
discussed previously.

Several observations can be made from this example.  First, the absence
of a mass pole in the vacuum Green's function of a colored object is
sufficient for a description of confinement provided the subsequent
interactions with other colored objects contain a scale beyond which
the interaction diminishes allowing the presence of the vacuum to
become apparent.  This defines a confinement scale that is largely
responsible for the characteristics of bound states. The effect of such
a scale is investigated in the numerical work to follow in Sec.4,
however, one can anticipate that as the range of the interaction
increases, the density of bound-state poles increases; finally giving
rise to a continuous spectrum as illustrated above.  Second, the
identification of a true bound-state pole requires careful examination
of the $P^2-q^2$ plane.  Finally, with regard to electron scattering,
in general the singularity structure of the QPV in the time-like region
of photon momentum squared dominates the behavior in the low momentum
region of space-like $q^2$.  This feature is exhibited here despite the
absence of a bound-state pole.  The intercept at $q^2=0$, is provided
by the constraints, (\ref{2.10}), derived from the WI, which the
coefficient functions given in (\ref{3.4}) satisfy explicitly.  Further
out in the space-like region  a transition to asymptotic behavior is
expected.  As is illustrated in Fig.3b, an unphysically sharp
transition occurs here due to the nonanalytic structure of the
self-energy amplitude, $B$, in (\ref{3.2}).  Based on these
observations, the approximation (\ref{3.1}) does not provide a useful
description of bound states or the approach to asymptopia, but is of
pedagogical value for investigating the behavior of the QPV for the
situation where the confinement length scale, defined here as the quark
separation at which the vacuum becomes apparent, is taken to infinity.

\section{Numerical evaluation}

\subsection{Methods and approximations}

The set of equations for the five transverse coefficient functions
represented in (\ref{2.16}), and given explicitly in the appendix,
forms the basis for the numerical investigation undertaken here.  The
approach used in evaluating these equations is to express each {\it
element} of the matrices(vectors) in the five-component space of
coefficient functions as a {\it matrix}({\it vector}) in the
direct-product space of the magnitude of the momentum $P$ and the
direction cosine $C_{Pq}$.  The value of the photon momentum, $q^2$, is
a parameter of the equations.  That is, for each value of $q^2$ a
different set of equations is obtained.  For the numerical evaluation,
the momentum $P$ and the direction cosine $C_{Pq}$ are discretized in
terms of Gauss-quadrature points.  The integrations are then carried
out as matrix multiplication.  The coupled equations are  solved as a
matrix equation in the expanded direct-product space of the coefficient
functions and the Gauss integration points.  The dimension of the
equation is given by 5$\times $nppts$\times $ncpts, where nppts and
ncpts are the number of Gauss points representing the $P$ and $C_{Pq}$
integrations respectively.  The number of points in both integrations
is varied to ensure accuracy.
For the momentum, twenty to forty quadrature points are used, while six
to ten are used for the direction cosine.

\subsubsection{Gluon two-point function ansatz}

The aim in this initial investigation is not necessarily to reproduce
the known bound-state spectrum, but rather to correlate trends in the
spectrum with changes in the parameterization of the infrared form of
the gluon two-point function.  The sensitivity of the spectrum to these
changes then provides a constraint on the calculation of the QPV in the
space-like region.  To this end,  a sufficient parametrization of the
gluon two-point function is given by
\begin{equation}
g^2D(P^2)=\frac{3\pi ^4\eta ^2}{16}R_0^4e^{-R_0^2P^2/4}+\frac{16\pi
^2}{11P^2\mbox{ln}\left( e+P^2/\Lambda ^2\right) },\label{4.1}
\end{equation}
where the first term (which has the coordinate space representation
$e^{-r^2/R_0^2}$) provides a length scale, $R_0$, characterizing the
infrared behavior while the second term models the known ultraviolet
form.  The parameter $\Lambda $, which is associated with the QCD
scale, is fixed at $1$ fm$^{-1}$ throughout this investigation.  As
$R_0$ approaches infinity the first term reduces to the delta-function
model of Eq.(\ref{3.1}).  The gluon two-point function (\ref{4.1}) has
been employed previously in the study of Bethe-Salpeter and
Schwinger-Dyson equations\cite{PCR89}, and is expected to provide an
acceptable accounting of the infrared effects for the purposes of the
investigation conducted here.

\subsubsection{Quark self energy ansatz}

The solution of the Bethe-Salpeter equation in the time-like region
requires knowledge of the quark Green's functions, and hence the
self-energy functions, for complex values of their arguments.  The
solution of the Schwinger-Dyson equation in the complex plane is a
challenging problem in itself\cite{burden92,conchi}, and is being
pursued elsewhere for forms such as that in
(\ref{4.1})\cite{burdencon}.  Although these solutions are crucial to a
comprehensive study of the present problem, reasonable approximations
are available.  The analytic solutions given in (\ref{3.2}) for the
self-energy functions obtained using the simple model two-point
function (\ref{3.1}) are useful in this regard.  It is shown, for
example, in Ref.\cite{PCR89} that the solutions to (\ref{2.2}) and
(\ref{2.3}) for the self-energy amplitudes obtained using (\ref{4.1}),
display behavior similar to the simple model solutions of (\ref{3.2})
on the real axis with the exception of an additional tail on the scalar
amplitude, $B$, attributed to the ultraviolet contribution in
(\ref{4.1}).  Others have argued on both theoretical and experimental
grounds that the asymptotic form of the scalar contribution to the self
energy should behave as $B(P^2)\sim 4m^3_D/P^2$, which, in the Landau
gauge, is exact within a logarithm\cite{pagels79}.  The self-energy
functions used in the numerical work here are thus modeled as
\begin{eqnarray}
A(P^{2}) & = & \left\{
\begin{array}{cl}
2, & P^{2}\leq \frac{\eta ^{2}}{4} \\
\frac{1}{2}\left[ 1+\left(1+\frac{2\eta ^{2}}{P^{2}}\right)
^{\frac{1}{2}}\right] , & P^{2}\geq \frac{\eta ^{2}}{4}
\end{array}\right. \nonumber \\ & &  \nonumber \\
B(P^{2}) & = & \left\{
\begin{array}{cl}
(\eta ^{2}-4P^{2})^{\frac{1}{2}}, &
\; \; \; \; \; \; \; \; \; \; \; \; P^{2}\leq \frac{\eta ^{2}}{6} \\
\frac{4m_D^3}{P^2}, &  \; \; \; \; \; \; \; \; \; \; \; P^{2}\geq
\frac{\eta ^{2}}{6}
\end{array}\right. , \label{4.2}
\end{eqnarray}
where the quantity $m_D$ has been referred to previously as the
``dynamical-quark mass"\cite{pagels79,cahill85}, which sets the scale
for dynamical chiral symmetry breaking.  Its value is determined here
by demanding that the function $B$ and its first derivative match at
the point $\eta ^2/6$, and is thus given by
$m_D=[24\sqrt{3}]^{-1/3}\eta \approx 0.29\eta $.  For $\eta \sim 1$
GeV, a dynamical-quark mass of $m_D\sim 290$ MeV is obtained, which is
consistent with the notion of a constituent-quark mass.  In
Ref.\cite{pagels79}, using evidence from $e^+e^-$ annihilation, the
dynamical mass is estimated to be $m_D\approx 244$ MeV.  These
arguments place constraints on the range of the parameter $\eta $,
which is thereby fixed at $5$ fm$^{-1}$ ($\approx 1$ GeV) for the
remainder of this work.  The ansatz (\ref{4.2}) encompasses the
expected behavior of the solutions to the Schwinger-Dyson equation
(\ref{SDE}) on the real axis\cite{conchi}, and is here taken to define
the solutions in the portion of the complex plane sampled by the QPV
calculation.

\subsection{Results}
\subsubsection{Time-like $q^2$ and bound states}

Shown in Fig.\ref{fig5} are the solutions to Eq.(\ref{2.16}) for the
five transverse coefficient functions in the time-like region of photon
momentum squared.  Two different values of the infrared scale, $R_0$,
are reported, from which it is observed that the spectrum is quite
sensitive to changes in this parameter.  In particular, an increase in
$R_0$ leads to an increase in the density of bound-state poles.  This
situation is, for example, analogous to a potential well, where $R_0$
plays the role of the range while the barrier is provided by the
repulsive interaction with the vacuum.  With the value of $R_0$ set to
$1$ fm, the ground-state mass occurs at $880$ MeV.  This is in
reasonable agreement with the observed $\rho $ mass, considering that
pion dressing is expected to lower this value\cite{rhomass}, yet the
excitation spectrum does not coincide with  experimentally observed
quantities.  Comparison with the measured excitation spectrum can be
improved by decreasing the range, $R_0$, however this also leads to an
increase in the ground-state mass, and away from the known value.  As
illustrated previously with the analytic solutions (\ref{3.4}), in the
limit as $R_0$ approaches infinity a continuous spectrum is obtained.
{}From this discussion, and the behavior displayed in Fig.\ref{fig5}, it
is evident that the combination of the gluon two-point function
(\ref{4.1}) and the self-energy functions (\ref{4.2}), is not
sufficient to reproduce the empirical excitation spectrum.
Nevertheless, it is encouraging that this approach affords the
investigation of spectra beyond the ground state, which until
recently\cite{MJ92} has been the
 focus of Bethe-Salpeter phenomenology applied to mesons.  The
reproduction of spectra is clearly essential to constraining the
interaction between quarks.  In this regard, the utility of the
approach presented here is that, given a gluon two-point function and
the associated solution to the Schwinger-Dyson equation, the vector
bound-state spectrum can be obtained.  The ability to reproduce the
experimentally observed spectrum is then contingent upon the capacity
of the ladder approximation to capture the relevant dynamics of QCD.

The extraction of bound-state wavefunctions is achieved by considering
the dependence of the coefficient functions, $\lambda ^T_i$, on the
momentum $P^2$ and the direction cosine $C_{Pq}$ in the vicinity of a
pole.  Shown in Fig.\ref{fig6} are the solutions for the ground-state
with mass $880$ MeV, and with $R_0=1$ fm, as a function of $P^2$ for
the two limiting  values of the direction cosine, $C_{Pq}=0,1$.  The
normalization is chosen such that the largest coefficient function is
unity at $P^2=C_{Pq}=0$.  The relative strength of the other
coefficients can then be observed.  The absolute normalization is
obtained through application of Eq.(\ref{2.21}), but is not relevant to
the present discussion.
A rough estimate for the dependence of the solutions on the momentum
and the direction cosine can be obtained by fitting the curves to an
exponential of the form $e^{-P^2(1+aC_{Pq}^2)/\Delta ^2}$, which gives
$a\sim 0.29$ and $\Delta \sim 3.5$ fm$^{-1}$.  The corresponding
estimate of the spatial extent is $r_V\sim 2/\Delta \sim 0.57$ fm.

\subsubsection{Space-like $q^2$ and electron scattering}

The description of the QPV in the space-like region is largely
independent of the spectrum in  the time-like region with the exception
of the ground state, which dominates the low momentum structure.  The
approach to asymptotic behavior at higher momentum is primarily
determined by the asymptotic forms of the gluon two-point function and
the self-energy functions.  Care has been taken here to accommodate
both of these low and high momentum attributes in a manner that is
consistent with their known features.  The results of the calculation
in the space-like region are summarized in Figs.\ref{fig7} and \ref{fig8}.

In Fig.\ref{fig7} the solutions for the coefficient functions, $\lambda
^T_i$, are displayed versus the photon momentum, $q^2$, for
$P^2=C_{Pq}=0$.  A smooth transition from the low momentum region,
dominated by the ground state vector-meson pole, to the high momentum
approach to asymptopia is observed.  It is therefore expected that the
range of applicability of these solutions spans the entire range of
momentum transfers available in electron scattering experiments.  With
the exception of $\lambda ^T_3$, which approaches unity asymptotically,
the coefficient functions vanish for large momenta consistent with
asymptotic freedom\cite{marciano78}.  The asymptotic form is apparent
at $q^2\sim 50$ fm$^{-2}$ ($\sim 2$ GeV$^2$), where the coefficient
function $\lambda ^T_3$ achieves roughly twice the strength of the
other coefficients.  It is also worth pointing out that to a very good
approximation the coefficients $\lambda ^T_2$ and $\lambda ^T_5$ are
equal over the momentum range reported.

Shown in Fig.\ref{fig8} are the coefficient functions versus the
momentum $P^2$ for $q^2=10$ fm$^{-2}$, and for the limiting values of
the direction cosine, $C_{Pq}=0,1$.  Only a modest angular dependence
is apparent, and again the coefficients $\lambda ^T_2$ and $\lambda
^T_5$ are approximately equal.  The observed behavior in the space-like
region can be roughly characterized in terms of the scales of the
problem by the following analytic forms:
\begin{eqnarray}
\lambda ^T_i&\approx &a_i\frac{1}{1+q^2/M^2_{\rho
}}\frac{1}{1+P^2/(4m_D^2)}\left[ \mbox{ln}\left( e+(P^2+q^2/4)/\Lambda
^2\right) \right] ^{-1/2}\; \; \; i\neq 3\nonumber \\
\lambda ^T_3&\approx &1+\frac{1}{1+q^2/M^2_{\rho }}\left[
\mbox{ln}\left( e+(P^2+q^2/4)/\Lambda ^2\right) \right] ^{-1/2},\label{4.3}
\end{eqnarray}
where the constants $a_i$ can be read from Fig.\ref{fig7}, and in
present case the ground state vector mass is $M_{\rho }=880$ MeV.  The
falloff with the momentum $P^2$ is reminiscent of the self-energy
functions, while that with $q^2$ reflects the presence of the
singularity in the time-like region.  It should be understood that
these forms are not the result of a detailed fit, but are rather
offered as a guide to the qualitative behavior over the momentum range
reported.

\section{Summary}
The inhomogeneous Bethe-Salpeter equation in the ladder approximation
has been used to study several issues that arise in the nonperturbative
description of the electromagnetic interaction with quarks.  The
investigation presented here represents a first solution of this
equation in a gauge invariant formulation which features confinement
and asymptotic freedom, in addition to the dynamical generation of
$q\bar{q}$ vector meson modes.  The principle results are summarized in
Figs.\ref{fig5}--\ref{fig8}.

The solution in the time-like region warrants further study,
particularly with regard to the reproduction of the excitation
spectrum.  Nevertheless, the description presented here shows promise
in the ability to study the vector meson spectrum beyond the ground
state.  Further, the demonstrated sensitivity of the spectrum to the
introduced infrared scale has positive implications for constraining
the largely unknown behavior of the gluon two-point function at low
momentum.  The utility of this approach is, however contingent upon the
ability of the ladder approximation to capture the relevant bound-state
dynamics of QCD.

The solution in the space-like region is potentially the most useful
result of this investigation.  The quark-based study of the EM
properties of hadrons relies heavily on the capacity to describe the
quark-photon interaction.  Modeling this quantity based on gauge
invariance provides only partial constraints on its behavior, and in
particular allows for arbitrariness in the components transverse to the
four momentum of the photon, which are those probed by electron
scattering experiments.  The use of vector meson dominance or the bare
QPV, is limited by the momentum range for which these approximations
are sensible.  Here a somewhat more fundamental approach has been
taken by describing the quark-photon interaction through a model in
which the dynamics dictate the role of $q\bar{q}$ vector bound states
and the asymptotic behavior.  The transverse contributions to the QPV
are thus uniquely defined within the employed ladder approximation to
the inhomogeneous Bethe-Salpeter equation, and no restrictions on the
momentum range are implied.  The solutions obtained here are therefore
expected to provide an essential ingredient in the quark-based
description of the EM interactions with hadrons.


\acknowledgements

This work was supported by the Department of Energy under Grant No.
DE-FG06-90ER40561.  The author wishes to thank C.D. Roberts, P.C.
Tandy, M. Burkardt and M. Herrmann for helpful conversations.

\unletteredappendix{Details of Eq.(\ref{2.16})}

The essential steps required to reduce the inhomogeneous Bethe-Salpeter
equation (\ref{IBS}) to the matrix form given in (\ref{2.16}) are
listed below.  To begin, the coefficient functions are isolated using
(\ref{2.6}) and (\ref{2.8}), and the orthogonality of the Dirac matrices to
obtain
\begin{eqnarray}
\frac{(P^T)^2}{\eta ^2}\lambda ^T_1&=&\frac{1}{4}\frac{P^T_{\mu }}{\eta
}\mbox{tr}\left[ \Gamma _{\mu }(P,q)\right] \nonumber \\
\frac{(P^T)^2}{\eta ^2}\lambda ^T_2-3\lambda ^T_3&=&-\frac{i}{4}\left(
\delta _{\mu \nu }-\frac{q_{\mu }q_{\nu }}{q^2}\right) \mbox{tr}\left[
\gamma _{\nu }\Gamma _{\mu }(P,q)\right] \nonumber \\
\frac{(P^T)^2}{\eta ^2}\lambda ^T_2-\lambda
^T_3&=&-\frac{i}{4}\frac{P^T_{\mu }P^T_{\nu }}{(P^T)^2} \mbox{tr}\left[
\gamma _{\nu }\Gamma _{\mu }(P,q)\right] \label{2.11} \\
\frac{P^T\cdot q}{\eta ^2}\lambda ^T_2+\frac{q^2P\cdot q}{\eta
^4}\lambda ^T_4&=&-\frac{i}{4}\frac{P^T_{\mu }q_{\nu }}{(P^T)^2}
\mbox{tr}\left[ \gamma _{\nu }\Gamma _{\mu }(P,q)\right] \nonumber \\
\frac{q^2(P^T)^2}{\eta ^4}\lambda ^T_5&=&-\frac{i}{8}\epsilon _{\mu \nu
\alpha \beta }\frac{P_{\alpha }q_{\beta }}{\eta ^2}\mbox{tr}\left[
\gamma _{\nu }\gamma _5\Gamma _{\mu }(P,q)\right] \nonumber .
\end{eqnarray}
The right-hand side of the equations in (\ref{2.11}) are evaluated
using the inhomogeneous Bethe-Salpeter equation (\ref{IBS}) for $\Gamma _{\mu
}$.

The remaining obstacle is the evaluation of the four-dimensional
integration present in (\ref{IBS}), which, upon substitution into
(\ref{2.11}), produces expressions of the form
\begin{eqnarray}
{\cal I}\equiv \int d^4K & &\hat{D}\left( P^2\! +\! K^2\! -\! 2PK\!
\sqrt{1-C^2_{Pq}}C_{KT}-\! 2PKC_{Pq}C_{Kq}\right) \nonumber \\
& &\; \; \; \; \; \; \; \; \times C^n_{KT}
F\left( P^2,K^2,q^2,C_{Kq},C_{Pq}\right) ,\label{2.12}
\end{eqnarray}
where $n=0,1,2$; $C_{Kq}$($C_{KT}$) is the direction cosine between $K$
and $q$($K$ and $P^T$); and $\hat{D}\left( (P-K)^2\right) \equiv
g^2D\left( (P-K)^2\right)/\left( 3\eta ^2\pi ^4\right )$.  Here $F$ is
a generic function representing the integrands obtained from
(\ref{2.11}).  Since the vectors $P^T_{\mu }$ and $q_{\mu }$ are
orthogonal, they can be used to define two of the four axes of
integration.  Rewriting the integration as
\begin{equation}
\int d^4K=2\int_{0}^{\infty }dK K^3\int
_{-1}^{1}dC_{Kq}dC_{KT}dC_1dC_2\delta
(1-C^2_{Kq}-C^2_{KT}-C^2_1-C^2_2),\label{2.13}
\end{equation}
where $C_1$ and $C_2$ are the remaining direction cosines, then obtains
from (\ref{2.12}) the result
\begin{equation}
{\cal I}=2\pi \int_{0}^{\infty }dK K^3\int _{-1}^{1}dC_{Kq}F\left(
P^2,K^2,q^2,C_{Kq},C_{Pq}\right) \hat{D}_n(P^2,K^2,C_{Pq},C_{Kq}),\label{2.14}
\end{equation}
where
\begin{equation}
\hat{D}_n( P^2,K^2,C_{Pq},C_{Kq})\equiv \int
_{-x}^{x}dC_{KT}\hat{D}\left( P^2\! +\! K^2\! -\! 2PK\!
\sqrt{1-C^2_{Pq}}C_{KT}-\! 2PKC_{Pq}C_{Kq}\right) C^n_{KT},\label{2.15}
\end{equation}
and $x\equiv \sqrt{1-C_{Kq}^2}$.

The explicit form of the quantities in (\ref{2.16}) is listed below.
The vector $b$ is given by
\begin{equation}
b=(0,3,1,0,0).\label{a1}
\end{equation}
The matrix ${\cal N}$ is given by
\begin{equation}
\begin{array}{ll}
{\cal N}_{11}=1&
{\cal N}_{22}=-\frac{(P^T)^2}{\eta ^2}\\
{\cal N}_{23}=3&
{\cal N}_{32}=-\frac{(P^T)^2}{\eta ^2}\\
{\cal N}_{33}=1&
{\cal N}_{42}=1\\
{\cal N}_{44}=\frac{q^2}{\eta ^2}&
{\cal N}_{55}=1,
\end{array}
\end{equation}
where all other elements are zero.  The matrix ${\cal M}$ is given by
\begin{equation}
\begin{array}{ll}
{\cal M}_{11}=2\pi \hat{D}_1F_0\frac{K}{P^T}&
{\cal M}_{12}=2\pi \hat{D}_1\frac{K^3}{P^T\eta
^2}\left(V+\frac{q^2}{\eta ^2}C_{Kq}^2W\right) \\
{\cal M}_{13}=-2\pi \hat{D}_1\frac{K}{P^T}V&
{\cal M}_{14}=2\pi \hat{D}_1\frac{K^3q^^2}{P^T\eta
^4}C_{Kq}^2\left(V+\frac{q^2}{\eta ^2}W\right) \\
{\cal M}_{21}=-\pi \hat{D}_0\frac{K^2}{\eta ^2}x^2V&
{\cal M}_{22}=\pi \hat{D}_0\frac{K^2}{\eta ^2}x^2\left(
F_1-2\frac{K^2}{\eta ^2}T\right) \\
{\cal M}_{23}=-\pi \hat{D}_0\left( 3F_1-2\frac{K^2}{\eta ^2}T\right) &
{\cal M}_{24}=-2\pi \hat{D}_0\frac{K^4q^2}{\eta ^6}x^2C_{Kq}^2T\\
{\cal M}_{25}=2\pi \hat{D}_0\frac{K^2q^2}{\eta ^4}x^2T&
{\cal M}_{31}=-\pi \hat{D}_2\frac{K^2}{\eta ^2}V\\
{\cal M}_{32}=\pi \hat{D}_2\frac{K^2}{\eta ^2}\left(
F_1-2\frac{K^2}{\eta ^2}T\right) &
{\cal M}_{33}=-\pi \left( F_1\hat{D}_0-2\frac{K^2}{\eta ^2}T\hat{D}_2\right) \\
{\cal M}_{34}=-2\pi \hat{D}_2\frac{K^4q^2}{\eta ^6}C_{Kq}^2T&
{\cal M}_{35}=\pi \left( \hat{D}_0x^2-\hat{D}_2\right) \frac{K^2q^2}{\eta
^4}T\\
{\cal M}_{41}=\pi \hat{D}_1\frac{K^2}{P^LP^T}C_{Kq}\left(
V+\frac{q^2}{\eta ^2}W\right) &
{\cal M}_{42}=-\pi \hat{D}_1\frac{K^2}{P^LP^T}C_{Kq}F_2\\
{\cal M}_{43}=-2\pi \hat{D}_1\frac{K^2}{P^LP^T}C_{Kq}T&
{\cal M}_{44}=-\pi \hat{D}_1\frac{K^2}{P^LP^T}C_{Kq}\frac{q^2}{\eta ^2}F_3\\
{\cal M}_{53}=-\pi \hat{D}_1\frac{K}{P^T}T&
{\cal M}_{55}=\pi \hat{D}_1\frac{K}{P^T}F_0
\end{array}
\end{equation}
where all unlisted elements are zero, and
\begin{equation}
\begin{array}{ll}
F_0\equiv \left( \frac{q^2}{4}-K^2\right) \frac{1}{\eta ^2}T+U&
F_1\equiv \left( K^2-\frac{q^2}{4}\right) \frac{1}{\eta ^2}T+U\\
F_2\equiv F_1+2\left( \frac{q^2}{4}-K^2\right) \frac{1}{\eta ^2}&
F_3\equiv F_1+2\left( \frac{q^2}{4}-K^2C_{Kq}^2\right) \frac{1}{\eta ^2}\\
T\equiv \eta ^4\alpha (K^2_{+})\alpha (K^2_{-})&
U\equiv \eta ^2\beta (K^2_{+})\beta (K^2_{-})\\
V\equiv \eta ^3\left[ \alpha (K_{+})\beta (K_{-})+\alpha (K_{-})\beta
(K_{+})\right] &
W\equiv \eta ^5 \left[ \frac{\alpha (K_{+})\beta (K_{-})-\alpha
(K_{-})\beta (K_{+})}{2K\cdot q}\right]
\end{array}
\end{equation}
The quantities $\alpha $ and $\beta $ are defined in (\ref{2.4});
$T,U,V,$ and $W$ are dimensionless functions of $K^2,q^2$ and
$C_{Kq}^2$; and the components of the momentum $P$ are defined as
$P_L\equiv PC_{Pq}$ and $P_T\equiv P\sqrt{1-C^2_{Pq}}$.

\figure{The EM vertex for a composite pion at tree level.  The quark
Green's functions, QPV, and the pion Bethe-Salpeter amplitudes, are
dressed in a consistent, gauge-invariance manner.\label{fig1}}

\figure{The ladder approximation for (a) the QPV and (b) the quark self
energy.  The Green's functions are defined by their inverse,
$G^{-1}(P)=i\! \not \! P+\Sigma (P)$.\label{fig2} }

\figure{The solutions for the five transverse coefficient functions,
$\lambda _i^T$, obtained using the simple gluon two-point function
described by Eq.(\protect{\ref{3.1}}) in the text, are plotted as a
function of the photon momentum $q^2$ in both the (a) time-like and (b)
space-like regions for $P^2=C_{Pq}=0$.\label{fig3} }

\figure{The coefficient function $\lambda _3^T$ is plotted in the
$q^2-P^2$ plane for $C_{Pq}=0$.  The variation of the singularity with
$P^2$ is indicative of a continuous spectrum.\label{fig4}}

\figure{The five transverse coefficient functions are plotted versus
time-like photon momentum $q^2$ for $P^2=C_{Pq}=0$.  The dependence
upon the infrared scale, $R_0$, is illustrated by comparing (a) where
$R_0=1$ fm with (b) where $R_0=0.75$ fm.\label{fig5}}

\figure{The five transverse coefficient functions are plotted versus
the momentum $P^2$ for values of the direction cosine (a) $C_{Pq}=0$
and (b) $C_{Pq}=1$, and for $-q^2=M_V^2=(880$ MeV$)^2$.\label{fig6}}

\figure{The five transverse coefficient functions are plotted versus
the photon momentum $q^2$ for $P^2=C_{Pq}=0$\label{fig7}}

\figure{The five transverse coefficient functions are plotted versus
the momentum $P^2$ for photon momentum $q^2=10$ fm$^{-2}$, and the
limiting values of the direction cosine (a) $C_{Pq}=0$ and (b)
$C_{Pq}=1$.\label{fig8}}


\begin{references}
\bibitem[1]{fnt1}
The Euclidean metric, in which $a\cdot a=a_{\mu }a_{\mu }$ and $\{
\gamma _{\mu },\gamma _{\nu }\} = 2\delta _{\mu \nu }$, is used
throughout this work.
\end{references}
\end{document}